\documentclass[]{aa}

\usepackage{amsmath,amssymb,amsfonts}
\usepackage{array,tabularx,multirow}
\usepackage{physics}
\usepackage{bm}
\usepackage{graphicx}
\usepackage{natbib}

\usepackage{xcolor}
\usepackage[normalem]{ulem}
\usepackage{cancel}

\usepackage{txfonts}

\begin{document}

\title{Occultation polarimetry as a probe of unresolved polarization}

%Solar and stellar occultation polarimetry
%Occultation polarimetry of unresolved radio sources

\titlerunning{Occultation polarimetry}

\author{A. G. Tevzadze\inst{1,2}}

\institute{
Evgeni Kharadze Georgian National Astrophysical Observatory,
Abastumani, Georgia
\and
Department of Physics, Faculty of Exact and Natural Sciences,
Ivane Javakhishvili Tbilisi State University, Tbilisi, Georgia
}

%\date{Received --; accepted --}

\abstract
% Context
{Low polarization in an unresolved source may reflect
weak local polarization or spatial cancellation between strongly polarized
regions. An occultation can break this spatial degeneracy by
differentially sampling the source as the occulting boundary moves.}
% Aims
{We develop a general theoretical framework for occultation
polarimetry and determine what information can
be inferred directly from unresolved full Stokes light curves,
without reconstructing a polarization image or prescribing a source
model.}
% Methods
{For a stationary, spatially incoherent source undergoing monotonic 
occultation, we relate Stokes variations to the polarization sampled 
by an arbitrarily curved occulting boundary and derive an exact 
variance decomposition of the normalized Stokes vector.}
% Results
{The polarization deficit decomposes exactly into three 
non-negative contributions: local polarization deficit, 
mixing within limb samples, and mixing between their mean 
polarization states. The last is directly measurable and provides 
a sharp, model-independent lower bound on total spatial mixing. 
The associated covariance tensor further characterizes the direction 
and correlations of the resolved polarization variations, 
including those between linear and circular polarization. 
Finite sampling preserves the bound, which can be strengthened 
through complementary occultation trajectories.}
% Conclusions
{Occultation polarimetry provides a powerful observational technique 
for extracting quantitative, model independent constraints on 
spatial polarization mixing directly from unresolved Stokes light curves. 
It offers a means to reveal both the extent and the form of polarization 
structure otherwise inaccessible at the available angular resolution.}

          
\maketitle

\section{Introduction}
\label{sec:intro}

Polarization is among the key diagnostics of astrophysical radiation,
offering insight into emission mechanisms, magnetic geometry, 
and propagation effects in sources ranging from the solar atmosphere 
to distant galaxies.

When emitting structures are spatially unresolved, contributions with 
different polarization orientations or opposite handedness can partially 
or completely cancel. A low observed polarization fraction may therefore 
indicate weak local polarization or cancellation among strongly polarized 
regions within the observing beam.
This ambiguity arises across astrophysical sources and wavelengths and 
is well documented in radio astronomy, where both spatial cancellation
and Faraday depolarization can reduce the observed polarization \citep{Burn1966}. 
Recent observations further illustrate how unresolved spatial structure 
and angular resolution affect the measured polarization \citep{DeRubeis2024}.
The distinction between mean polarization and its second moments
also underlies statistical diagnostics of depolarization
\citep{MelroseMacquart1998}. 

As a foreground body moves across a background source, the resulting
occultation can reveal spatial variations in its brightness and 
polarization beyond the angular resolution of the instrument.
As a moving boundary scans the
source, classical occultation techniques recover one-dimensional
brightness information on angular scales finer than the nominal
instrumental resolution \citep{Scheuer1965}.
Eclipse mapping techniques also reconstruct spatial brightness
distributions from unresolved light curves through maximum entropy
inversion \citep{Horne1985}. During a solar eclipse, the lunar limb
provides the scanning boundary. LOFAR observations of lunar
de-occultation of the solar radio corona have revealed structure
below the resolution of standard interferometric images
\citep{Ryan2021}. In transiting systems, photometric light curve
slopes have also been used to constrain the geometry and motion of
occulting edges \citep{vanWerkhoven2014,vanDam2024}.

Occultations also provide information on spatially varying
polarization. Their polarimetric signatures have been investigated
in circumstellar scattering envelopes
\citep{BrownFox1989,FoxBrown1991}, in exoplanet transits and related
occultation geometries
\citep{CarciofiMagalhaes2005,Veenstra2025,Michaelis2025}, and in
X-ray eclipses of active galactic nuclei \citep{MarinDovciak2015}.

Direct evidence that occultations can reveal fine polarization
structure comes from radio observations of the 1963 lunar occultation
of the Crab Nebula, which detected variations in linear polarization
across the source \citep{Davies1966}. More recently,
\citet{MilesPaez2019} reported enhanced linear polarization during
planetary transits of TRAPPIST-1, providing observational support
for transit induced polarization. Modeling studies have examined
the polarization evolution of the Crab Nebula during lunar occultation
\citep{NakamuraShibata2007} and of $\epsilon$ Aurigae in and out of
eclipse \citep{Ignace2021}. 
Recent IXPE observations of eclipsing X-ray binaries have likewise
revealed significant polarization changes associated with eclipse,
interpreted as changes in the relative contributions of spatially
distinct emitting or scattering regions
\citep{Ninoyu2025,Mikusincova2026,Anitra2026}. These results
highlight the spatial selectivity and diagnostic potential of
occultation polarimetry.

Complementary inverse approaches recover stellar limb polarization
from unresolved photometric and polarimetric eclipse light curves
through regularized inversion \citep{Coleman1998}. Such approaches
interpret the measurements through specified source geometries,
polarization models, or inversion assumptions.

A broader question is what can be inferred about spatial polarization 
mixing directly from occultation measurements, without reconstructing 
a polarization map or prescribing a source model.

To address this question, we develop a general theoretical
framework for unresolved full Stokes occultation measurements
of a stationary source during monotonic ingress or egress. The
framework relates changes in the observed coherency matrix
to the polarization of emission sampled by a moving occulting
boundary. An exact variance identity for the normalized
Stokes vector separates the polarization deficit into three
non-negative contributions: local polarization deficit, mixing within
each boundary sample, and mixing between samples. Unlike
occultation imaging or source specific polarization modeling,
the present diagnostic yields a model independent quantitative
lower bound on spatial polarization mixing directly from unresolved
Stokes light curves. The associated polarization covariance further
retains directional information, distinguishing different forms of
linear and circular polarization variation that are lost in the
scalar mixing measure.

The inferred lower bound on spatial polarization mixing is sharp,
meaning that the observed light curves are consistent with 
no larger spatial contribution. The remaining polarization deficit 
may arise from weak local polarization, mixing within limb samples, 
or both. A non-zero bound therefore establishes the presence of 
spatial polarization mixing, whereas a vanishing bound does not 
exclude it. The diagnostic constrains spatial mixing in the 
observed radiation, regardless of whether the polarization 
differences originate in the emission or in propagation.

Section 2 develops the 
coherency matrix formulation of polarimetric occultations 
and relates unresolved Stokes variations to emission sampled 
by the moving boundary. Section 3 derives the exact decomposition 
of the polarization deficit, establishes a sharp lower bound 
on spatial polarization mixing, and introduces directional 
covariance diagnostics. Section 4 develops observational estimators, 
examines finite sampling and characteristic polarization signatures,
and illustrates the method through synthetic and 
complementary occultation scans. 
Section 5 discusses the observational implications and limitations 
of the framework and summarizes the main conclusions.

\section{Polarimetric occultation formalism}
\label{sec:occultation}

Let $\bm{r}=(x,y)$ denote angular coordinates in the local tangent plane
of the sky, neglecting curvature over the field of interest.
In a fixed orthonormal linear polarization basis, the local polarization
state can be described by the brightness coherency matrix
\citep{TevzadzeShatashvili2026,BornWolf1999,Hamaker1996} as follows:
\begin{equation}
{\bf J}(\bm{r}) = \frac{1}{2}
\left(\begin{array}{cc}
I(\bm{r})+Q(\bm{r}) & U(\bm{r})-\mathrm{i}V(\bm{r}) \\
U(\bm{r})+\mathrm{i}V(\bm{r}) & I(\bm{r})-Q(\bm{r})
\end{array}\right) ~.
\label{eq:Jlocal}
\end{equation}
Here, $I(\bm{r}), Q(\bm{r}), U(\bm{r})$, and $V(\bm{r})$ are the local
Stokes brightness parameters. The parameter $I(\bm{r})$ describes the
total intensity, $Q(\bm{r})$ and $U(\bm{r})$ specify the 
linear polarization, and $V(\bm{r})$ specifies
the signed circular polarization, with its sign convention fixed by
Eq.~(\ref{eq:Jlocal}). The diagonal elements give the intensities in the
two basis components, while the off-diagonal elements describe their
mutual correlation. The Stokes parameters may represent either a single
frequency channel or a specified observing band.
Hereafter, explicit spatial dependence is omitted where unambiguous, 
while frequency dependence is suppressed except when discussed explicitly.

The coherency matrix ${\bf J}$ is positive semidefinite, so the
intensity measured in any polarization state is non-negative. We define
the polarized Stokes vector,  polarization vector for $I>0$ and 
local degree of polarization as
\begin{equation}
\bm S \equiv
\begin{pmatrix}
Q\\ U\\ V
\end{pmatrix} ~,
\qquad
\bm P \equiv \frac{\bm S}{I} ~,
\qquad
\eta=|\bm P| ~.
\label{eq:plocal}
\end{equation}
Hereafter, all expressions involving division by an intensity are
understood to apply only where that intensity is positive.
The local degree of polarization can be expressed as
\begin{equation}
\eta^2 = \frac{|\bm S|^2}{I^2} = 1-\frac{4\det{\bf J}}
{({\rm Tr}\,{\bf J})^2} ~.
\label{eq:etaJ}
\end{equation}
For the matrix in Eq.~(\ref{eq:Jlocal}),
\begin{equation}
{\rm Tr}\,{\bf J}=I ~,
\qquad
\det{\bf J}=\frac14\left(I^2-|\bm S|^2\right)\geq 0 ~.
\end{equation}
At $\eta=0$, the radiation is unpolarized and
${\bf J}=(I/2)\mathbb{I}_2$, where $\mathbb{I}_2$ is the
$2\times2$ identity matrix. At $\eta=1$, it is fully polarized
and ${\bf J}$ has rank one, with $\det{\bf J}=0$.
Intermediate values $0<\eta<1$ describe partially polarized
radiation.

Let the opaque occulting boundary be defined implicitly by
$\mathcal{L}(x,y,t)=0$, with $\mathcal{L}>0$ in the visible
region and $\mathcal{L}<0$ in the occulted region.
This description accommodates an arbitrarily curved boundary,
including the lunar limb. We assume that the source brightness and
polarization remain stationary during the scan, emission from
distinct sky positions is mutually incoherent, and diffraction
is negligible.

\begin{figure}
\centering
\includegraphics[width=0.98\columnwidth]{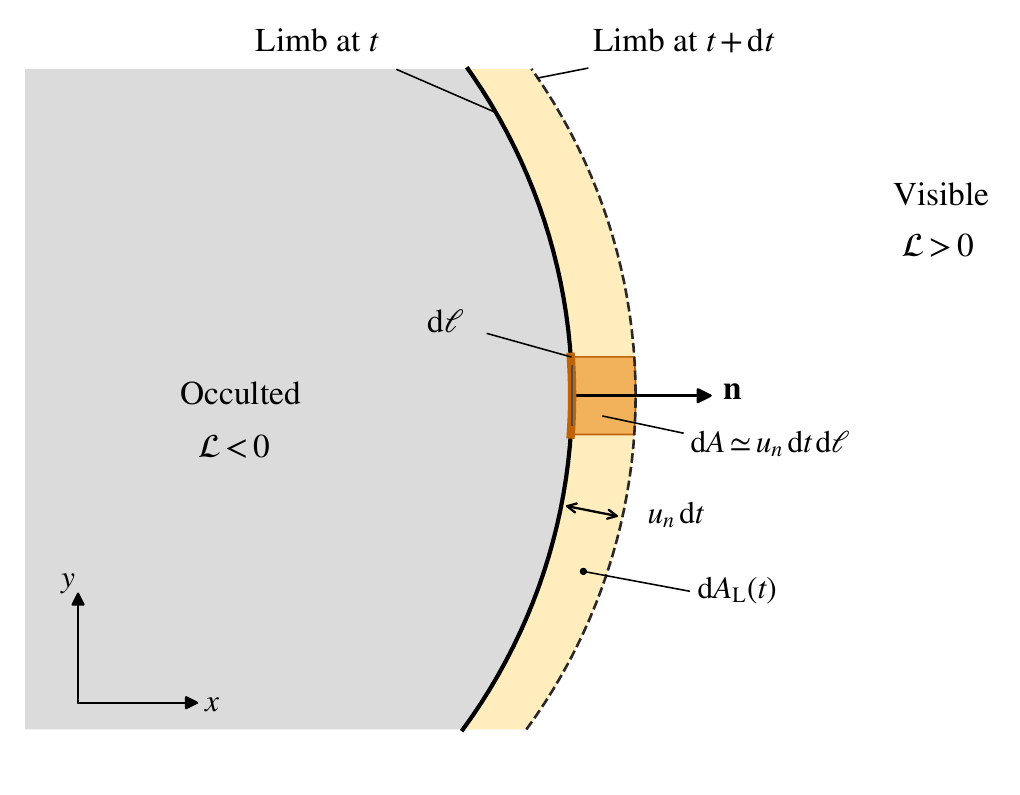}
\caption{Schematic geometry of a curved occulting limb during
ingress. The solid and dashed curves indicate the limb positions
at $t$ and $t+\mathrm{d}t$, respectively. A limb element of
arc length $\mathrm{d}\ell$ sweeps an angular area
$\mathrm{d}A=u_n\,\mathrm{d}t\,\mathrm{d}\ell$
to first order (dark shading), where $u_n\,\mathrm{d}t$
is the local normal displacement, which varies along the limb.
The pale band represents the total swept area
$\mathrm{d}A_{\rm L}(t)$ across the source.}
\label{fig:limb_geometry}
\end{figure}

On the occulting boundary $\mathcal L(\bm r,t)=0$, where
$|\nabla\mathcal L|>0$, the unit normal pointing from the
occulted to the visible region is
$\bm n=\nabla\mathcal L/|\nabla\mathcal L|$, with
$\nabla=(\partial_x,\partial_y)$. 
As a point moves with the boundary by $\mathrm{d}\bm r$
during $\mathrm{d}t$, it remains on $\mathcal L=0$,
implying $\mathrm{d}\mathcal L=0$ and hence
\begin{equation}
\nabla\mathcal L\cdot\mathrm{d}\bm r
+\partial_t\mathcal L\,\mathrm{d}t=0 ~.
\label{eq:limb_motion}
\end{equation}
Its velocity normal to the boundary is therefore
\begin{equation}
u_n =
\bm n\cdot\frac{\mathrm{d}\bm r}{\mathrm{d}t} =
-\frac{\partial_t\mathcal L}{|\nabla\mathcal L|} ~.
\label{eq:vn}
\end{equation}
With this orientation, $u_n>0$ denotes local ingress as the
limb advances into the visible region, while $u_n<0$ denotes
local egress. During pure ingress or pure egress, we define
the occultation sign parameter as
\begin{equation}
\sigma =
\begin{cases}
+1, & \text{ingress} ~,\\
-1, & \text{egress} ~.
\end{cases}
\end{equation}
Since $\sigma u_n=|u_n|$ wherever the limb crosses the source,
the positive area swept by an element $\mathrm{d}\ell$ during
$\mathrm{d}t$ is
$\mathrm{d}A=\sigma\,u_n\,\mathrm{d}t\,\mathrm{d}\ell$.
Integrating the local area element along the source intersecting
limb defines the swept angular area $\mathrm{d}A_{\rm L}(t)$:
\begin{equation}
\mathrm{d}A_{\rm L}(t) =
\sigma\,\mathrm{d}t \int_{\mathcal L(\bm r,t)=0} u_n\,\mathrm{d}\ell ~.
\label{eq:dAL}
\end{equation}
Figure~\ref{fig:limb_geometry} illustrates the area
$\mathrm{d}A_{\rm L}(t)$ swept by the occulting limb during
$\mathrm{d}t$. Its local width is the normal displacement
$|u_n|\,\mathrm{d}t$, which generally varies along a curved limb.

Assuming uniform instrumental weighting across the source, 
the observed coherency matrix of the unresolved emission is
\begin{equation}
{\bf J}_{\rm obs}(t) = \int_{\mathcal L(\bm r,t)>0}
{\bf J}(\bm r)\,\mathrm{d}^2\bm r ~.
\label{eq:Jobs}
\end{equation}
The observed intensity and polarized Stokes vector are denoted by 
$I_{\rm obs}(t)$ and $\bm S_{\rm obs}(t)$, respectively, 
and are obtained by integrating the corresponding local quantities 
over the visible region.

As the limb moves, an element of arc length $\mathrm{d}\ell$
changes the observed coherency matrix by 
$-u_n\mathbf J(\bm r)\,\mathrm{d}t\,\mathrm{d}\ell$ to first order in $\mathrm{d}t$.
Integrating along the source intersecting limb gives
\begin{equation}
\frac{\mathrm{d}{\bf J}_{\rm obs}}{\mathrm{d}t} =
-\int_{\mathcal L(\bm r,t)=0}
u_n{\bf J}(\bm r)\,\mathrm{d}\ell ~.
\label{eq:stripJ}
\end{equation}
Although Eq.~(\ref{eq:Jobs}) is a 2D integral, its time
variation is determined entirely by the 1D moving boundary and
therefore reduces to an integral along the limb element
$\mathrm{d}\ell$.
The four observed Stokes light curves obey this relation componentwise.
Note that if the limb simultaneously covers and uncovers different regions, 
the observed change combines two contributions that must be separated 
before applying the present formalism.

To distinguish between unresolved, local and occultation limb-sampled
polarization, we use three polarization degrees throughout
the analysis:
\begin{equation}
\begin{aligned}
\eta_{\rm obs} &= \eta_\star &:& \quad \text{unresolved source degree} \,,\\
\eta &= \eta(\bm r) &:& \quad \text{local degree} \,,\\
\eta_{\rm L} &= \eta_{\rm L}(t)
&:& \quad \text{limb-sampled degree} \,.
\end{aligned}
\label{eq:polarization_hierarchy}
\end{equation}
Here, $\eta_{\star}$ denotes the integrated degree of polarization 
of the unresolved source,
$\eta(\bm r)$ is the local degree of polarization that varies across
the source, and $\eta_{\rm L}(t)$ is the degree of polarization of the
combined emission from the area $\mathrm{d}A_{\rm L}(t)$ swept by
the moving limb across the source.
These polarization degrees are independent of the choice of basis.
Before integration, however, the local Stokes vectors must be expressed
in a common basis to preserve their relative polarization states and any
resulting cancellation.

We now take $\mathcal R$ to be the source region swept by the limb
during a monotonic occultation interval
$[t_{\rm init},t_{\rm fin}]$ of pure ingress or egress, with each
point crossed exactly once. 
For a local quantity $F(\bm r)$, its intensity weighted mean over
the scanned region is
\begin{equation}
\langle F\rangle =
{\displaystyle\int_{\mathcal R} I(\bm r)F(\bm r)\,\mathrm{d}^2\bm r}
\,\Big/\, {\displaystyle\int_{\mathcal R} I(\bm r)\,\mathrm{d}^2\bm r} ~.
\label{eq:scan_average}
\end{equation}
The corresponding limb-sampled mean is
\begin{equation}
\langle F\rangle_{\rm L}(t) =
{\displaystyle\int_{\mathcal L=0}
u_n I(\bm r)F(\bm r)\,\mathrm{d}\ell}
\,\Big/\, {\displaystyle\int_{\mathcal L=0} u_n I(\bm r)\,\mathrm{d}\ell} ~.
\label{eq:strip_average}
\end{equation}
The factor $u_n$ weights each limb element in proportion to the
angular area it sweeps per unit time; its common sign cancels in
the ratio.

The mean polarization vector and degree of polarization of the
scanned region are
\begin{equation}
\bm P_{\rm scan} = \langle\bm P\rangle ~,
\qquad 
\eta_{\rm scan} = |\bm P_{\rm scan}| ~,
\label{eq:Pscan}
\end{equation}
while the corresponding limb-sampled quantities are
\begin{equation}
\bm P_{\rm L}(t) = \langle\bm P\rangle_{\rm L}(t) ~,
\qquad \eta_{\rm L}(t) = |\bm P_{\rm L}(t)| ~. 
\label{eq:PL}
\end{equation}

For a short interval $\mathrm{d}t$ of pure ingress or egress, the
moving limb removes or restores the emission sampled during that
interval. The corresponding differential coherency matrix is
\begin{equation}
\mathrm{d}{\bf J}_{\rm L} = -\sigma\,
\frac{\mathrm{d}{\bf J}_{\rm obs}}{\mathrm{d}t}\,
\mathrm{d}t ~.
\label{eq:dJL}
\end{equation}
Thus, $\eta_{\rm L}(t)$ is the polarization degree of the combined
emission from the area $\mathrm{d}A_{\rm L}(t)$ swept by the limb.
It includes cancellation among polarization states within that area
and therefore need not equal the mean local polarization degree.
Applying Eq.~(\ref{eq:etaJ}) to $\mathrm{d}{\bf J}_{\rm L}$ gives
\begin{equation}
\eta_{\rm L}^2(t) =
1- \frac{4\det\!\left(\mathrm{d}{\bf J}_{\rm obs}/\mathrm{d}t\right)}
{\left[ \mathrm{Tr}\!\left(\mathrm{d}{\bf J}_{\rm obs}/\mathrm{d}t\right)
\right]^2} ~.
\label{eq:pstripJ}
\end{equation}
Equations~(\ref{eq:dJL})--(\ref{eq:pstripJ}) require no image
reconstruction. During pure ingress or egress, derivatives of the
unresolved full Stokes signal determine the mean polarization state
of the emission sampled by the moving limb, independently of the
limb shape. Polarization variation within each limb sample remains
unresolved.

Second order moments of the Stokes parameters contain information on
spatial polarization variation that is not captured by their mean
values alone.
Related covariance based descriptions have been used to
characterize polarization fluctuations and depolarization in radio
polarimetry
\citep{MelroseMacquart1998,vanStraten2009,McKinnon2009}.
Following this general approach, we introduce a covariance tensor of
the polarization vectors to retain information beyond the scalar
measure of polarization mixing. Thus we define
\begin{equation}
C_{ij} = \Big\langle P_iP_j\Big\rangle
- \Big\langle P_i\Big\rangle \Big\langle P_j\Big\rangle ~,
\label{eq:Cmix}
\end{equation}
where $i,j\in\{Q,U,V\}$. 
The components of $\bm C$ describe the directional structure and
correlations of polarization variation in Stokes space. 
Therewith, the covariance components satisfy the Cauchy--Schwarz inequality
\begin{equation}
C_{ij}^{2}\leq C_{ii}C_{jj} ~.
\label{eq:C_corrbound}
\end{equation}
The corresponding scalar measure is given by the trace,
\begin{equation}
{\rm Tr}\,\bm C =
\left\langle\eta^2\right\rangle -
\left|\Big\langle\bm P\Big\rangle\right|^2 =
\Big\langle\eta^2\Big\rangle - \eta_{\rm scan}^2 ~.
\label{eq:Cmix_trace}
\end{equation}
Thus, ${\rm Tr}\,\bm C$ quantifies the total spatial polarization mixing 
in quadratic form, while $\bm C$ retains its directional structure 
and correlations between normalized Stokes components.

\begin{figure}
\centering
\includegraphics[width=0.98\columnwidth]{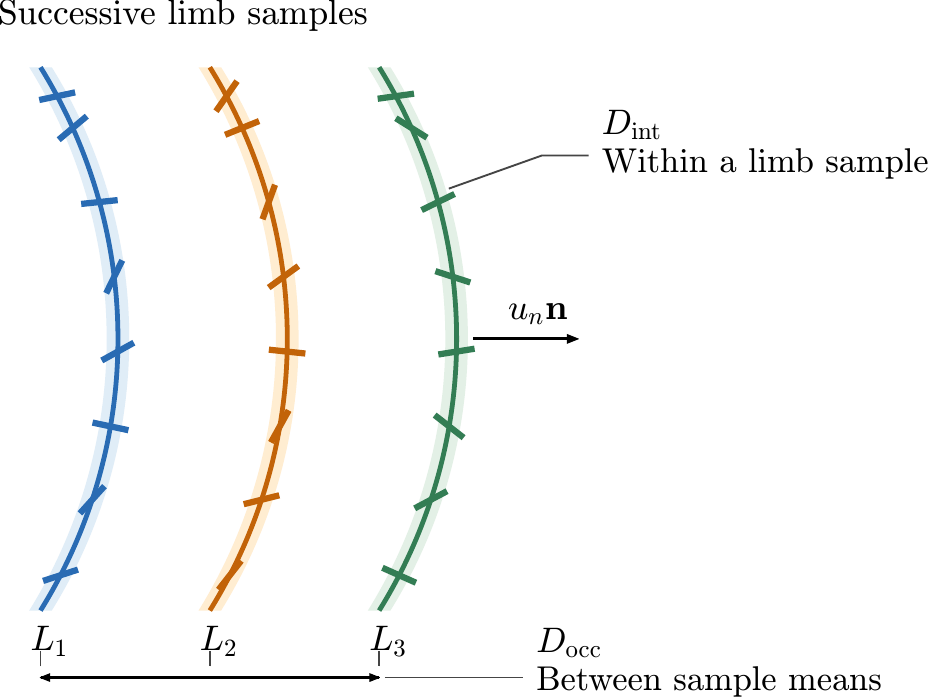}
\caption{Schematic illustration of spatial polarization mixing during an occultation. 
The curved bands represent successive limb samples, while the short strokes indicate 
local linear polarization orientations. Variations within individual samples 
contribute to $D_{\rm int}$, whereas differences between their mean polarization 
states contribute to $D_{\rm occ}$. The arrow indicates the local normal velocity 
$u_n\bm n$ of the occulting limb.}
\label{fig:polarization-mixing}
\end{figure}

\section{Decomposition of the polarization deficit}
\label{sec:depol}

The polarization deficit of an unresolved source can reflect 
incomplete local polarization or the spatial mixing of different 
polarization states. A single unresolved measurement cannot separate 
these contributions, but an occultation can resolve part of the spatial mixing.
We therefore derive an exact decomposition of the polarization deficit 
and identify the contribution that can be measured directly 
from occultation observations.

\subsection{Scalar decomposition and occultation bound}
\label{sec:scalar_decomposition}

To quantify the contribution of spatial polarization mixing, 
we first decompose the polarization deficit of the region swept 
by the occulting limb. For a complete scan, $\mathcal R$ is the 
entire source and $\eta_{\rm scan}=\eta_{\rm obs}$, whereas 
for a partial scan, the analysis is restricted to the 
scanned region $\mathcal R$. We define its quadratic 
polarization deficit as
\begin{equation}
D_{\rm scan} \equiv 1-\eta_{\rm scan}^{2} ~.
\label{eq:deficit_scan}
\end{equation}
The local contribution is the intensity weighted mean local
polarization deficit,
\begin{equation}
D_{\rm loc} \equiv
\left\langle 1-\eta^{2}\right\rangle =
1-\left\langle\eta^{2}\right\rangle ~.
\label{eq:Dloc}
\end{equation}
From Eq.~(\ref{eq:Cmix_trace}), the total spatial polarization mixing
within the scanned region is
\begin{equation}
D_{\rm mix} \equiv
{\rm Tr}\,\bm C = \left\langle\eta^{2}\right\rangle
- \eta_{\rm scan}^{2} ~.
\label{eq:Dmix}
\end{equation}
The polarization deficit can therefore be decomposed into 
local and spatial contributions,
\begin{equation}
D_{\rm scan} = D_{\rm loc} + D_{\rm mix} ~.
\label{eq:deficit_local_mix}
\end{equation}

An occultation further separates the spatial contribution according
to the polarization structure resolved by successive limb samples.
These samples contribute to the scanned region in proportion to their
sampled intensity. The corresponding intensity per unit time is
\begin{equation}
I_{\rm L}(t) \equiv
-\sigma\,\frac{\mathrm{d}I_{\rm obs}}{\mathrm{d}t}
= \sigma\int_{\mathcal L=0}
u_n I(\bm r)\,\mathrm{d}\ell ~,
\label{eq:IL}
\end{equation}
which defines the normalized occultation weight
\begin{equation}
W(t) \equiv
{I_{\rm L}(t)} \, \Big/ \,
{\displaystyle\int_{t_{\rm init}}^{t_{\rm fin}}
I_{\rm L}(\tau)\,\mathrm{d}\tau} ~.
\label{eq:W}
\end{equation}

During a monotonic scan, successive limb samples exhaust the scanned
emission, so their intensity weighted mean polarization equals the
scan-averaged polarization,
\begin{equation}
\bm P_{\rm scan} = \int_{t_{\rm init}}^{t_{\rm fin}}
W(t)\,\bm P_{\rm L}(t)\,\mathrm{d}t ~,
\label{eq:Pscan_strip}
\end{equation}
while
\begin{equation}
\left\langle\eta^2\right\rangle =
\int_{t_{\rm init}}^{t_{\rm fin}} W(t) \,
\left\langle\eta^2\right\rangle_{\rm L}(t) \,
\mathrm{d}t ~.
\label{eq:eta2_strip}
\end{equation}

We now evaluate the covariance of the successive limb-sampled 
polarization states to quantify the spatial mixing resolved by the occultation,
\begin{equation}
C^{\rm occ}_{ij} \equiv 
\int_{t_{\rm init}}^{t_{\rm fin}} W(t)\,
P_{{\rm L},i}(t)P_{{\rm L},j}(t)\, \mathrm{d}t -
P_{{\rm scan},i}P_{{\rm scan},j} ~.
\label{eq:Cocc}
\end{equation}
Its trace gives the occultation resolved contribution to the spatial
mixing,
\begin{equation}
D_{\rm occ} \equiv
{\rm Tr}\,\bm C^{\rm occ} =
\int_{t_{\rm init}}^{t_{\rm fin}}
W(t)\eta_{\rm L}^{2}(t)\,\mathrm{d}t -
\eta_{\rm scan}^{2} ~.
\label{eq:Docc}
\end{equation}
The remaining internal spatial mixing occurs within individual limb samples,
\begin{equation}
D_{\rm int} \equiv
\int_{t_{\rm init}}^{t_{\rm fin}} W(t)
\left[ \left\langle\eta^2\right\rangle_{\rm L}(t) -
\eta_{\rm L}^{2}(t) \right] \,\mathrm{d}t ~.
\label{eq:Dint}
\end{equation}
Thus, Eqs. (\ref{eq:eta2_strip})--(\ref{eq:Dint}) lead to the following:
\begin{equation}
D_{\rm mix} = D_{\rm occ}+D_{\rm int} ~,
\label{eq:Dmix_decomp}
\end{equation}
and therefore
\begin{equation}
0 \leq D_{\rm occ}\leq D_{\rm mix} ~.
\label{eq:Docc_bound}
\end{equation}
The lower bound is sharp and attainable. Assigning each position
the polarization state sampled at its limb crossing time
reproduces the observed Stokes variations with $D_{\rm int}=0$.
Thus, the data are consistent with $D_{\rm mix}=D_{\rm occ}$,
and no stronger lower bound can be established from a single
scan without additional assumptions.
Hence, $D_{\rm occ}$ provides a direct lower bound on the total spatial
polarization mixing. Combining Eq.~(\ref{eq:Dmix_decomp}) with
Eq.~(\ref{eq:deficit_local_mix}) gives the exact decomposition
\begin{equation}
D_{\rm scan} = D_{\rm loc}+D_{\rm occ}+D_{\rm int} ~.
\label{eq:deficit_total}
\end{equation}
The distinction between the two spatial mixing contributions is
illustrated in Fig.~\ref{fig:polarization-mixing}.

The quantity $D_{\rm occ}$ measures spatial polarization variation
resolved through differences between the mean states of successive
limb samples. It vanishes when all samples have the same mean
polarization state, even if polarization structure remains unresolved
within individual samples. A non-zero $D_{\rm occ}$ therefore directly
demonstrates spatial polarization variation across the scanned region.
Since $W(t)$ and $\bm P_{\rm L}(t)$ are determined from the unresolved
Stokes occultation curves, both $\bm C_{\rm occ}$ and $D_{\rm occ}$
are directly measurable without reconstructing a polarization map.

\subsection{Directional polarization diagnostics}
\label{sec:directional}

We now use the full covariance tensor to characterize the directional 
structure of polarization variations resolved by the occultation. 
Its diagonal elements quantify the variances of the normalized 
Stokes components, while the off-diagonal elements describe 
their correlations.

For non-zero mean linear polarization, the covariance can be resolved
relative to the mean linear polarization direction. Let us define
\begin{equation}
L_{\rm scan} =
\left( P_{{\rm scan},Q}^{2}+P_{{\rm scan},U}^{2} \right)^{1/2} ~,
\end{equation}
and, an orthonormal basis in the \(Q\)–\(U\) plane,
\begin{equation}
\bm e_{\parallel} = \frac{1}{L_{\rm scan}}
\begin{pmatrix}
P_{{\rm scan},Q}\\
P_{{\rm scan},U}
\end{pmatrix} ~,
\qquad
\bm e_{\perp} = \frac{1}{L_{\rm scan}}
\begin{pmatrix}
-P_{{\rm scan},U}\\
P_{{\rm scan},Q}
\end{pmatrix} ~,
\label{eq:lin_basis}
\end{equation}
where $\bm e_{\parallel}$ is parallel to the mean linear polarization
vector in the $Q$--$U$ plane and $\bm e_{\perp}$ is perpendicular to it.
The $Q$--$U$ covariance block is then defined as
\begin{equation}
\bm C^{\rm occ}_{\rm lin} =
\begin{pmatrix}
C^{\rm occ}_{QQ} & C^{\rm occ}_{QU}\\
C^{\rm occ}_{QU} & C^{\rm occ}_{UU}
\end{pmatrix} ~,
\end{equation}
with projected components
\begin{equation}
C^{\rm occ}_{ab} = \bm e_a\cdot
\left( \bm C^{\rm occ}_{\rm lin}\bm e_b \right) ~,
\label{eq:Cocc_linproj}
\end{equation}
where $a,b\in\{\parallel,\perp\}$ denote the directions
of the orthonormal basis $(\bm e_\parallel,\bm e_\perp)$.

To relate the projected covariance components to variations 
in linear polarization degree and position angle, 
we express the limb-sampled linear polarization as
\begin{equation}
P_{{\rm L},Q}+iP_{{\rm L},U} = L(t)e^{2i\chi(t)} ~.
\label{eq:linear_complex}
\end{equation}
Small variations along $\bm e_{\parallel}$ correspond predominantly
to changes in linear polarization degree, while those along
$\bm e_{\perp}$ correspond to polarization angle changes. Thus,
$C^{\rm occ}_{\parallel\parallel}$ and
$C^{\rm occ}_{\perp\perp}$ measure the corresponding resolved
variances, while $C^{\rm occ}_{\parallel\perp}$ measures their
covariance.

The linear--circular covariances can be projected onto the same basis.
Let us define
\begin{equation}
C^{\rm occ}_{aV} = e_{a,Q}C^{\rm occ}_{QV} + e_{a,U}C^{\rm occ}_{UV} ~,
\label{eq:Cocc_LVproj}
\end{equation}
Expanding Eq.~\eqref{eq:linear_complex} to first order
in variations of the linear polarization amplitude and
position angle about their mean values gives
\begin{align}
C^{\rm occ}_{\parallel V} &\simeq
\int W(t)
[L(t)-L_{\rm scan}] [P_{{\rm L},V}(t)-P_{{\rm scan},V}]
\,\mathrm{d}t ~,
\nonumber\\
C^{\rm occ}_{\perp V} &\simeq
2L_{\rm scan} \int W(t) [\chi(t)-\chi_{\rm scan}] 
[P_{{\rm L},V}(t)-P_{{\rm scan},V}] \,\mathrm{d}t ~,
\label{eq:Cocc_LVinterpret}
\end{align}
where $\chi_{\rm scan}$ is the position angle of the mean
linear polarization vector.

The off-diagonal terms of the polarization covariance matrix
distinguish three forms of co-variation:
\begin{itemize}
\item $C^{\rm occ}_{\parallel\perp}$ ~:~ polarization degree--angle covariance;
\item $C^{\rm occ}_{\parallel V}$ ~:~ polarization degree--circular covariance;
\item $C^{\rm occ}_{\perp V}$ ~:~ polarization angle--circular covariance.
\end{itemize}
These correlations reveal aspects of polarization structure beyond 
those captured by the scalar mixing measure.

Having determined the individual covariance components, we can construct 
a basis independent measure of the resolved linear--circular covariance,
\begin{equation}
C^{\rm occ}_{LV} \equiv
\left[ \left(C^{\rm occ}_{QV}\right)^2 +
\left(C^{\rm occ}_{UV}\right)^2 \right]^{1/2} =
\left[ \left(C^{\rm occ}_{\parallel V}\right)^2 +
\left(C^{\rm occ}_{\perp V}\right)^2 \right]^{1/2} ~.
\label{eq:CLV}
\end{equation}
When the denominator is non-zero, its normalized strength is
\begin{equation}
\rho_{LV} \equiv
\frac{C^{\rm occ}_{LV}}
{\left[ \left(C^{\rm occ}_{QQ}+C^{\rm occ}_{UU}\right)
C^{\rm occ}_{VV} \right]^{1/2}} ~.
\label{eq:rhoLV}
\end{equation}
Values near zero indicate weak linear--circular covariance, 
while values approaching unity indicate a strong correlation 
between circular-polarization variations and a preferred direction 
in the linear-polarization plane.
Importantly, such correlations constrain the underlying polarization 
structure but do not uniquely identify the responsible physical mechanism.

When the mean linear polarization vanishes, $L_{\rm scan}=0$, 
the parallel--perpendicular decomposition becomes undefined. 
Nevertheless, the directional structure of the $Q$--$U$ covariance 
can still be characterized by the basis independent anisotropy parameter,
\begin{equation}
A_L \equiv
\frac{ \left[ \left(C^{\rm occ}_{QQ}-C^{\rm occ}_{UU}\right)^2 +
4\left(C^{\rm occ}_{QU}\right)^2 \right]^{1/2}} 
{C^{\rm occ}_{QQ}+C^{\rm occ}_{UU}} ~,
\label{eq:AL}
\end{equation}
defined when $C^{\rm occ}_{QQ}+C^{\rm occ}_{UU}>0$.
Values near zero indicate nearly isotropic variation in the
$Q$--$U$ plane, whereas values approaching unity indicate variation
concentrated along a preferred direction. Thus, $A_L$ retains
directional information even when no mean linear polarization
direction exists.

The diagonal terms of the polarization covariance matrix
quantify variations along individual Stokes directions, while its 
off-diagonal terms reveal correlations between them. 
Such structure may reflect changes in polarized component balance, 
magnetic field geometry, or polarized transfer through a magnetized plasma.

Although these diagnostics do not uniquely identify the underlying 
physical mechanism, they do provide additional model-independent constraints. 
In particular, polarization patterns that produce the same scalar mixing 
can be distinguished by their anisotropy and covariance structure.

\section{Observational diagnostics and applications}
\label{sec:estimator}

The theoretical framework developed above can be applied directly 
to unresolved full Stokes observations. In this section, we derive 
practical estimators, examine the effects of finite sampling and 
frequency dependence, and illustrate the diagnostic potential of 
synthetic and complementary occultation scans. 
To facilitate the practical application of the framework, 
Table~\ref{tab:occultation_diagnostics} summarizes the 
required measurements, practical estimators, and 
physical information that can be inferred from them.

\subsection{Direct estimator and finite sampling}
\label{sec:binning}

For a monotonic interval of pure ingress or egress,
we define the signed Stokes increments
\begin{align}
\Delta I &=
\sigma\left[I_{\rm obs}(t_{\rm init}) -I_{\rm obs}(t_{\rm fin})\right] ~,
\nonumber\\
\Delta\bm S &= \sigma\left[\bm S_{\rm obs}(t_{\rm init})
-\bm S_{\rm obs}(t_{\rm fin})\right] ~.
\label{eq:Stokes_total_increment}
\end{align}
These determine the mean polarization state of the scanned emission,
\begin{equation}
\bm P_{\rm scan} = \frac{\Delta\bm S}{\Delta I} ~,
\qquad
\eta_{\rm scan}^{2} = \frac{|\Delta\bm S|^{2}}{(\Delta I)^{2}} ~.
\label{eq:Pscan_increment}
\end{equation}
Using Eq.~(\ref{eq:IL}), the occultation-resolved covariance
can be expressed directly in terms of the observed Stokes derivatives,
\begin{equation}
C^{\rm occ}_{ij} =
-\frac{\sigma}{\Delta I} \int_{t_{\rm init}}^{t_{\rm fin}}
\frac{ (\mathrm{d}S_{{\rm obs},i}/\mathrm{d}t) (\mathrm{d}S_{{\rm obs},j}/\mathrm{d}t)}
{\mathrm{d}I_{\rm obs}/\mathrm{d}t} \,\mathrm{d}t - P_{{\rm scan},i}P_{{\rm scan},j} ~,
\label{eq:Cocc_direct}
\end{equation}
where $i,j\in\{Q,U,V\}$. Its trace gives
$D_{\rm occ}={\rm Tr}\,\bm C^{\rm occ}$.
Thus, both scalar and directional diagnostics follow directly
from unresolved Stokes measurements, without polarization-image
reconstruction or knowledge of the occulting-boundary geometry.

For discretely sampled observations, the Stokes increments over 
successive intervals $[t_k,t_{k+1}]$ are given by
\begin{align}
\Delta I_k &=
\sigma\left[ I_{\rm obs}(t_k)-I_{\rm obs}(t_{k+1}) \right] ~,
\nonumber\\
\Delta\bm S_k &= 
\sigma\left[ \bm S_{\rm obs}(t_k)-\bm S_{\rm obs}(t_{k+1}) \right] ~.
\label{eq:Stokes_increments}
\end{align}
For bins with positive intensity increments, the corresponding 
mean polarization vectors and normalized intensity weights are
\begin{equation}
\bm P_k = \frac{\Delta\bm S_k}{\Delta I_k} ~,
\qquad
w_k = \frac{\Delta I_k}{\Delta I} ~,
\qquad
\sum_k w_k = 1 ~.
\label{eq:Pk_wk}
\end{equation}
The corresponding finite bin covariance is
\begin{equation}
C^{{\rm occ,bin}}_{ij} =
\sum_k w_k \left(P_{k,i}-P_{{\rm scan},i}\right)
\left(P_{k,j}-P_{{\rm scan},j}\right) ~,
\label{eq:Cocc_bin}
\end{equation}
where $\bm P_{\rm scan}=\sum_k w_k\bm P_k$.
Thus, its trace yields the occultation resolved mixing,
\begin{align}
D_{\rm occ}^{\rm bin} =
\sum_k w_k \left|\bm P_k-\bm P_{\rm scan}\right|^2
%\nonumber\\
= \frac{1}{\Delta I} \sum_k\frac{|\Delta\bm S_k|^2}{\Delta I_k}
- \frac{|\Delta\bm S|^2}{(\Delta I)^2} ~.
\label{eq:Docc_direct_discrete}
\end{align}
Bins with zero intensity increments carry no weight and are
omitted. For $D_{\rm scan}>0$, the fraction of the polarization
deficit resolved by the occultation is
\begin{equation}
f_{\rm occ}^{\rm bin} \equiv
\frac{D_{\rm occ}^{\rm bin}}{D_{\rm scan}}.
\label{eq:focc_bin}
\end{equation}
Since the bins form disjoint samples of the scanned emission,
the total spatial mixing separates into
\begin{equation}
D_{\rm mix} =
D_{\rm occ}^{\rm bin} + D_{\rm int}^{\rm bin} ~, 
\qquad
D_{\rm int}^{\rm bin}\geq0 ~,
\label{eq:Dmix_bin}
\end{equation}
which establishes the finite bin lower bound,
\begin{equation}
0\leq D_{\rm occ}^{\rm bin}\leq D_{\rm mix} ~.
\label{eq:bound_bin}
\end{equation}
Spatial mixing within individual bins remains unresolved
and contributes to $D_{\rm int}^{\rm bin}$, so finite sampling
preserves the lower bound character of the diagnostic. 
The polarization deficit not resolved by finite sampling
follows from Eq.~(\ref{eq:Dmix_bin}) as
\begin{equation}
D_{\rm scan}-D_{\rm occ}^{\rm bin} =
D_{\rm loc}+D_{\rm int}^{\rm bin} ~.
\label{eq:remaining_deficit}
\end{equation}

Measurement noise introduces uncertainty and bias through the
quadratic dependence on the Stokes increments and division by
the intensity increments, particularly when $\Delta I_k$ is small.
Correlations among Stokes measurements and between increments
sharing a common measurement must therefore be accounted for
\citep{Montier2015a,Montier2015b}.
Thus, the full measurement noise covariance matrix is essential
for reliable inference from noisy occultation data.

Refining a temporal partition resolves additional polarization
variation within the original bins. For nested partitions,
the resolved mixing satisfies
\begin{equation}
D_{\rm occ}^{({\rm fine})} =
D_{\rm occ}^{({\rm coarse})} + D_{\rm sub} ~,
\qquad
D_{\rm sub}\geq 0 ~,
\label{eq:nested}
\end{equation}
where $D_{\rm sub}$ is the additional mixing resolved by
subdividing the bins. Consequently,
\begin{equation}
D_{\rm occ}^{({\rm coarse})} \leq D_{\rm occ}^{({\rm fine})}
\leq D_{\rm mix} ~.
\label{eq:nested_bound}
\end{equation}
Successive refinement thus yields a monotonic hierarchy of 
increasingly strong lower bounds on spatial polarization mixing. 
The scalar mixing measure and basis independent covariance 
diagnostics also remain invariant under a spatially 
uniform rotation in the $Q$--$U$ plane.

\subsection{Physical signatures in the polarization covariance}
\label{sec:physical-signatures}

The covariance diagnostics can reveal organized polarization structure
that is weak or absent in the unresolved mean signal. Although they do
not uniquely identify the underlying physical mechanism, several
limiting cases can have characteristic signatures.

Particularly illustrative is the case of spatially varying Faraday rotation
\citep{Burn1966,Sokoloff1998}. 
For polarized emission characterized by a single rotation
measure at each sky position \citep{Brentjens2005},
the polarization angle follows
\begin{equation}
\chi(\bm r,\lambda) = \chi_0(\bm r)+{\rm RM}(\bm r)\lambda^2 ~.
\end{equation}
Here, $\chi_0(\bm r)$ is the polarization angle at zero
wavelength, and ${\rm RM}(\bm r)$ is proportional to the
line-of-sight integral of the product of thermal electron
density and the line-of-sight magnetic field component.
Spatial variations in ${\rm RM}$ can reduce the net polarization
of an unresolved source even when the emission remains strongly
polarized locally. When polarization-angle variations are small
and dominate over amplitude variations, the resolved fluctuations
are predominantly perpendicular to the mean linear-polarization
direction, with
\begin{equation}
C^{\rm occ}_{\perp\perp} \gg C^{\rm occ}_{\parallel\parallel} ~.
\end{equation}

For small polarization-angle variations and approximately
constant linear-polarization amplitude, the perpendicular
covariance is
\begin{equation}
C^{\rm occ}_{\perp\perp} \simeq 4L_{\rm scan}^{2}
\int W(t) [\chi(t)-\chi_{\rm scan}]^2\,\mathrm{d}t ~.
\end{equation}
When differential Faraday rotation dominates,
\begin{equation}
\chi(t)-\chi_{\rm scan}
\simeq [{\rm RM}_{\rm L}(t)-{\rm RM}_{\rm scan}]\lambda^2 ~,
\end{equation}
where ${\rm RM}_{\rm L}(t)$ is the effective rotation measure
of the limb sample and ${\rm RM}_{\rm scan}$ its weighted mean.
Consequently,
\begin{equation}
\frac{C^{\rm occ}_{\perp\perp}}
{4L_{\rm scan}^{2}} \simeq
\lambda^{4} \int W(t) [{\rm RM}_{\rm L}(t)-{\rm RM}_{\rm scan}]^2
\,\mathrm{d}t ~. 
\label{eq:faraday_covariance}
\end{equation}
If the intensity weights and effective rotation-measure
distribution are approximately wavelength independent,
Eq.~(\ref{eq:faraday_covariance}) predicts a
$\lambda^4$ scaling of the normalized perpendicular variance,
providing a characteristic signature of spatially varying
Faraday rotation in the small angle limit.

Another characteristic signature arises when the resolved
linear polarization variations are concentrated along a
preferred direction in the $Q$--$U$ plane. This may occur
when the relative strengths of competing polarization modes
vary across the source \citep{McKinnon2009}.
In the limiting 1D case, aligning the $Q$ axis
with this direction gives
\begin{equation}
C^{\rm occ}_{QU}=C^{\rm occ}_{UU}=0 ~,
\qquad A_L=1 ~,
\label{eq:anisotropy_limit}
\end{equation}
provided the linear polarization variance is non-zero.
By contrast, nearly isotropic variations yield $A_L\simeq0$.
Importantly, $A_L$ remains defined even when the mean linear
polarization vanishes, allowing organized polarization structure
to be detected in an apparently unpolarized source.

Correlations between linear and circular polarization provide 
an additional diagnostic. A dominant $C^{\rm occ}_{\parallel V}$ 
links circular polarization variations primarily to changes in 
linear polarization degree, whereas a dominant $C^{\rm occ}_{\perp V}$ 
links them to polarization angle variations in the small variation limit. 
Such correlations may arise from changes in the relative strengths of 
elliptically polarized modes or from polarized transfer involving 
generalized Faraday rotation and conversion \citep{KennettMelrose1998}. 
In particular, spatially varying mode balance can couple linear and 
circular polarization, while propagation-induced conversion 
can correlate changes in polarization angle with circular polarization. 
The normalized measure $\rho_{LV}$ quantifies the strength of the resolved 
linear--circular covariance.

Although these signatures do not uniquely identify the underlying 
physical processes, they provide complementary constraints on the 
geometry and spectral dependence of unresolved polarization structure, 
extending the information available from the scalar mixing measure alone.

\begin{figure*}[t]
\centering
\includegraphics[width=0.98\textwidth]{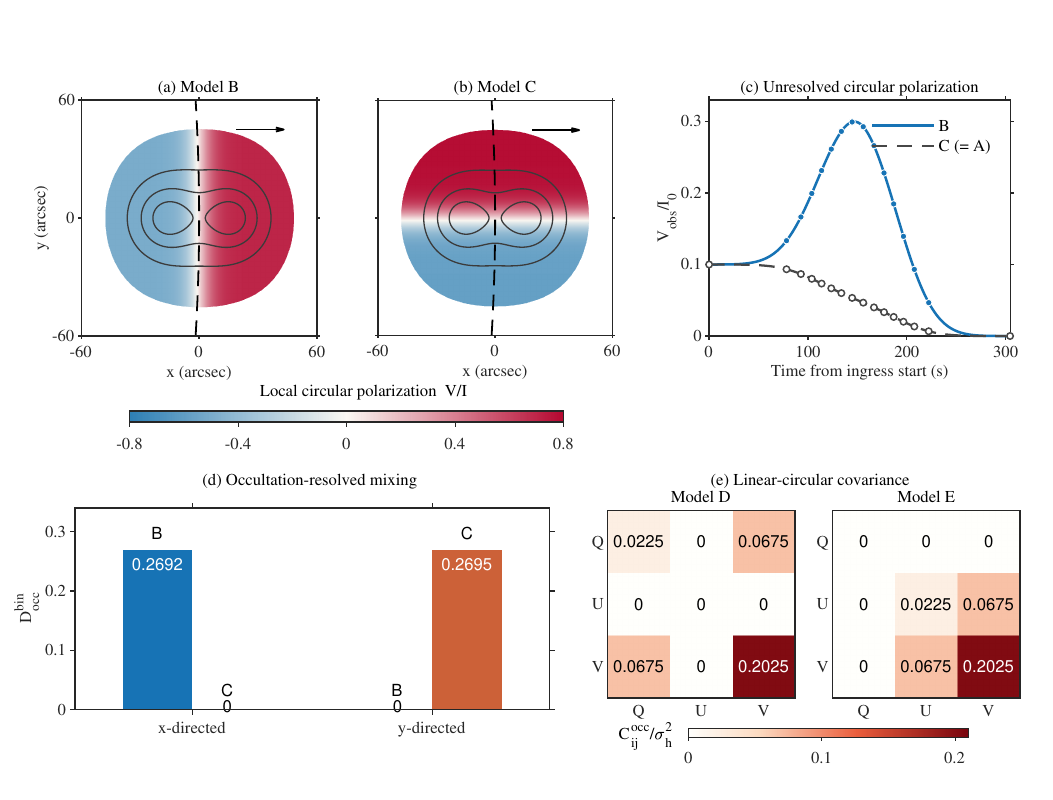}
\caption{Synthetic polarization maps, occultation signals,
and covariance diagnostics.
Panels a and b show the local $V/I$ of models B and C,
with intensity contours and the leading occulting limb
(dashed curves), whose motion is indicated by arrows.
Colour is suppressed in faint regions for clarity,
but all integrations include the full source.
Panel c shows $V_{\rm obs}/I_0$ during $x$-directed ingress,
with 16 snapshots defining 15 equal-intensity bins.
Although models B and C have identical total spatial mixing,
the $x$-directed scan resolves only B, while C remains
indistinguishable from the uniformly polarized model A.
Panel d compares the resolved mixing for orthogonal scans,
showing that the $y$-directed scan reverses this result.
Panel e shows the normalized covariance matrices
$C_{ij}^{\rm occ}/\sigma_h^2$ of models D and E,
demonstrating that identical scalar mixing can conceal
distinct linear--circular correlations.}
\label{fig:synthetic-occultation}
\end{figure*}

\subsection{Synthetic occultation and complementary scans}
\label{sec:synthetic-occultation}

We use a synthetic benchmark motivated by the structured circular
polarization of solar active regions
\citep{LeeWhite2011,Sych2023}. The brightness distribution,
common to all models, consists of two compact components
and a diffuse background,
\begin{equation}
I(x,y)=I_{\ast}T(x)T(y)
\left[G_{+}+G_{-}+bG_{\sigma_{\rm b}}(x,y)\right] ~,
\label{eq:synthetic-brightness}
\end{equation}
where
\begin{equation}
\begin{aligned}
G_{\pm} &= G_{\sigma}(x\pm d,y) ~,\\
G_{\sigma}(x,y) &= \exp\left[-\frac{x^2+y^2}{2\sigma^2}\right] ~,\\
T(z) &=
\begin{cases}
[1-(z/H)^2]^2 ~, & |z|<H ~,\\
0 ~, & |z|\geq H ~.
\end{cases}
\end{aligned}
\end{equation}
We adopt $H=60''$, $d=20''$, $\sigma=12''$,
$\sigma_{\rm b}=30''$, and $b=0.35$.
Three polarization patterns share this brightness distribution:
\begin{equation}
\frac{V}{I}=
\begin{cases}
p_0, & \mathrm{A},\\
p_0+a_x\tanh(x/\ell), & \mathrm{B},\\
p_0+a_y\tanh(y/\ell), & \mathrm{C},
\end{cases}
\label{eq:synthetic-polarization}
\end{equation}
with $Q=U=0$, $p_0=0.1$, $\ell=8''$, $a_x=0.6$, and
$a_y\simeq0.677011$, chosen so that B and C have equal spatial
mixing. All three models have an integrated polarization degree
of $0.1$, giving $D_{\rm scan}=0.99$. Model A has
$D_{\rm mix}=0$, whereas B and C both have
$D_{\rm mix}=0.2706$.

A circular occultor of radius $R_{\rm M}=1000''$ moves along
$x$ at $0.4''\,\mathrm{s}^{-1}$, producing complete ingress in
$304.5\,\mathrm{s}$. We calculate the diagnostics directly from
the unresolved Stokes increments using 15 bins of equal removed
intensity, $\Delta I_k=I_0/15$, where $I_0$ is the unobscured
flux. Figure~\ref{fig:synthetic-occultation} shows the
polarization maps and occultation diagnostics.

The adopted polarization scale of $8''$ corresponds to a
limb crossing time of approximately $20\,\mathrm{s}$,
requiring shorter sampling intervals and source stability
throughout the five minute ingress. For lunar occultations,
the Fresnel scale is a few arcseconds at decimetre wavelengths
but approaches the assumed spatial scale at metre wavelengths.
Observational applications therefore require adequate
temporal resolution, polarimetric sensitivity, and control
of diffraction and calibration effects.

For model B, the $x$-directed scan gives
$D_{\rm occ}^{\rm bin}=0.2692$, recovering $99.5\%$ of the true
spatial mixing. Increasing the temporal resolution approaches
the continuous value $D_{\rm occ}=0.2706$.

For model C, reflection symmetry about $y=0$ cancels the
spatially varying term in every limb sample, giving
$V_{\rm obs}(t)=0.1I_{\rm obs}(t)$. Its unresolved signal is
therefore identical to that of model A, and the $x$-directed
scan yields $D_{\rm occ}=0$ at any temporal resolution.
A complementary $y$-directed scan reverses this result:
model C gives $D_{\rm occ}^{\rm bin}=0.2695$, recovering
$99.6\%$ of the true mixing, while the mixing in B remains
unresolved.

For scans covering the same stationary region at a common
frequency and with consistent spatial weighting, the strongest
individual bound is
\begin{equation}
D_{\rm occ}^{\rm comp}
\equiv \max_a D_{\rm occ}^{(a)}
\leq D_{\rm mix},
\label{eq:complementary_bound}
\end{equation}
where $a$ labels the scan trajectories. Complementary scans
can thus reveal mixing hidden along individual trajectories.

The covariance tensor further distinguishes structures with
identical scalar mixing. Consider two additional polarization
patterns sharing the same brightness distribution, observed
along $x$. With $h(x)=\tanh(x/\ell)$, define
\begin{equation}
\begin{aligned}
\bm P_{\rm D} &=
(q_0+a_{\rm L}h,\,0,\,v_0+a_{\rm V}h),\\
\bm P_{\rm E} &=
(q_0,\,a_{\rm L}h,\,v_0+a_{\rm V}h),
\end{aligned}
\label{eq:synthetic-covariance}
\end{equation}
where $\bm P=(Q,U,V)/I$. We adopt
$(q_0,v_0,a_{\rm L},a_{\rm V})=(0.30,0.15,0.15,0.45)$,
satisfying $|\bm P|\leq1$.
Since the brightness is even in $x$, $\langle h\rangle=0$,
and the mean linear polarization lies along $Q$.
Defining the occultation-resolved variance
\begin{equation}
\sigma_h^2 \equiv
\int W(t)\langle h\rangle_{\rm L}^2(t)\,\mathrm{d}t,
\end{equation}
the projected linear--circular covariances become
\begin{align}
{\rm D}:&\quad
C_{\parallel V}^{\rm occ}
=a_{\rm L}a_{\rm V}\sigma_h^2,
\qquad C_{\perp V}^{\rm occ}=0,
\nonumber\\
{\rm E}:&\quad
C_{\parallel V}^{\rm occ}=0,
\qquad C_{\perp V}^{\rm occ}
=a_{\rm L}a_{\rm V}\sigma_h^2.
\label{eq:synthetic-covariance-result}
\end{align}
Yet both patterns give
\begin{equation}
D_{\rm occ}
=(a_{\rm L}^2+a_{\rm V}^2)\sigma_h^2,
\qquad \rho_{LV}=1.
\label{eq:synthetic-covariance-scalar}
\end{equation}
Thus, identical scalar mixing and linear--circular
correlation strength can conceal distinct covariance structures.
In the small-variation limit $|a_{\rm L}|\ll q_0$,
D represents predominantly degree--circular mixing,
whereas E represents angle--circular mixing.
Their projected cross terms distinguish the two.

\begin{table*}[]
\caption{Observational inputs and diagnostics of occultation
polarimetry. Stokes measurements are calibrated and unresolved.
All quantities refer to the region swept during a scan satisfying
the assumptions of the framework.}
\label{tab:occultation_diagnostics}
\centering
\begingroup
\small
\setlength{\tabcolsep}{4pt}
\renewcommand{\arraystretch}{1.15}

\begin{tabularx}{\textwidth}{
@{}
>{\raggedright\arraybackslash}p{0.18\textwidth}
>{\raggedright\arraybackslash}p{0.2\textwidth}
>{\raggedright\arraybackslash}p{0.14\textwidth}
>{\raggedright\arraybackslash}X
@{}}
\hline\hline
Observation &
Calculated quantity &
Reference &
Information inferred \\
\hline

Single frequency occultation scan\newline
$I_{\rm obs}(t_k)$,\,
$\bm S_{\rm obs}(t_k)$ &
\textit{Sample polarization}\newline
$\bm P_k=\Delta\bm S_k/\Delta I_k$ &
Eq. (\ref{eq:Pk_wk}) &
Mean polarization state of the emission removed or revealed
in each time bin. \\

\cline{2-4}
&
\textit{Scanned polarization}\newline
$\bm P_{\rm scan}$, $\eta_{\rm scan}$,\newline
$D_{\rm scan}=1-\eta_{\rm scan}^{2}$ &
Eq. (\ref{eq:Pscan_increment}) &
Net polarization and polarization deficit of the
swept emission. \\

\cline{2-4}
&
\textit{Polarization covariance}\newline
$\bm C^{\rm occ,bin}$ &
Eq. (\ref{eq:Cocc_bin}) &
Directional covariance of the polarization states resolved
between successive time bins. \\ 

\cline{2-4}
&
\textit{Linear polarization structure}\newline
$C^{\rm occ}_{\parallel\parallel}$,
$C^{\rm occ}_{\perp\perp}$,
$C^{\rm occ}_{\parallel\perp}$, $A_L$ &
Eqs. (\ref{eq:Cocc_linproj}), (\ref{eq:AL}) &
Distinguishes linear polarization degree and angle variations, with
$A_L$ measuring their $Q$--$U$ anisotropy. \\

\cline{2-4}
&
\textit{Linear--circular covariance}\newline
$C^{\rm occ}_{\parallel V}$,
$C^{\rm occ}_{\perp V}$,
$C^{\rm occ}_{LV}$, $\rho_{LV}$ &
Eqs. (\ref{eq:Cocc_LVproj} - \ref{eq:rhoLV}) &
Connects circular polarization changes with variations in linear degree
or angle, quantified by $C^{\rm occ}_{LV}$ and $\rho_{LV}$.\\

\cline{2-4}
&
\textit{Spatial mixing bound}\newline
$D_{\rm occ}^{\rm bin}
={\rm Tr}\,\bm C^{\rm occ,bin}$ &
Eqs. (\ref{eq:Docc_direct_discrete}),
(\ref{eq:bound_bin}) &
Model-independent lower bound on spatial polarization mixing required
by the observed light curves.\\

\cline{2-4}
&
\textit{Resolved deficit fraction}\newline
$f_{\rm occ}^{\rm bin}
=D_{\rm occ}^{\rm bin}/D_{\rm scan}$ &
Eq. (\ref{eq:focc_bin}) &
Fraction of the polarization deficit identified as spatial mixing.\\

\cline{2-4}
&
\textit{Remaining deficit}\newline
$D_{\rm scan}-D_{\rm occ}^{\rm bin}$ &
Eq. (\ref{eq:remaining_deficit}) &
Unresolved combination of local deficit and within bin spatial mixing.\\

\cline{2-4}
&
\textit{Sampling refinement}\newline
$D_{\rm occ}^{({\rm fine})}$, $D_{\rm sub}$ &
Eqs. (\ref{eq:nested} - \ref{eq:nested_bound}) &
Additional spatial mixing resolved by finer temporal sampling.\\

\hline

Frequency resolved occultation scan\newline
$I_{\rm obs}(t_k,\nu)$,\,
$\bm S_{\rm obs}(t_k,\nu)$ &
\textit{Frequency dependence}\newline
$\bm C^{\rm occ,bin}(\nu)$,
$D_{\rm occ}^{\rm bin}(\nu)$,\newline
$A_L(\nu)$, $\rho_{LV}(\nu)$ &
Eq. (\ref{eq:faraday_covariance}) &
Spectral evolution of the resolved polarization structure, including
signatures of spatially varying Faraday rotation. \\

\hline

Complementary occultation scans\newline
$I_{\rm obs}^{(a)}(t_k)$,\,
$\bm S_{\rm obs}^{(a)}(t_k)$ &
\textit{Complementary scan bound}\newline
$D_{\rm occ}^{\rm comp}$ &
Eq. (\ref{eq:complementary_bound}) &
Strongest individual scan bound, revealing spatial mixing hidden
along individual trajectories. \\

\hline
\end{tabularx}
\endgroup
\end{table*}

Finally, we assess noise sensitivity by fixing the 15 bin
boundaries and adding independent Gaussian errors with rms
$10^{-3}I_0$ to each Stokes measurement. For model B,
$2\times10^4$ realizations yield a standard deviation of
$2.2\times10^{-3}$ and a positive bias of $1.5\times10^{-3}$
in the raw $D_{\rm occ}^{\rm bin}$ estimator.
A noise only test with uniform model A gives a mean spurious
bound of $1.35\times10^{-3}$ and with 99\% of realizations
below $2.3\times10^{-3}$. Reliable observational bounds must
therefore account for measurement-noise and calibration
covariances.

\section{Discussion and conclusions}
\label{sec:discussion}

A low degree of polarization in an unresolved source can have
different origins. The emission may be weakly polarized locally, or
strongly polarized regions may cancel when their signals are combined.
As an occulting limb moves across the source, it changes the relative
contributions of these regions to the observed signal. We have
developed a framework in which unresolved full-Stokes light curves
yield the mean polarization state of the emission sampled successively
by the limb, providing spatial polarimetric information without
reconstructing a polarization map.

The main result is a directly measurable lower bound on the
contribution of spatial mixing to the polarization deficit. It follows
from unresolved Stokes variations without prescribing the source
morphology or the shape of the occulting boundary. A non-zero bound
establishes that spatial polarization variations contribute to the
observed depolarization, while the resolved fraction quantifies how
much of the deficit is accounted for by structure resolved during
the scan. The synthetic examples show that substantial spatial
mixing can be almost fully recovered along one trajectory while
remaining completely hidden along another.
The resulting observables provide complementary constraints on the
amount, direction, spectral dependence, and trajectory dependence
of the resolved polarization structure
(Table~\ref{tab:occultation_diagnostics}).

The bound is sharp: the same light curves cannot require a larger
spatial contribution from a single scan. Emission sampled
simultaneously along the limb remains unresolved, so the remaining
deficit may arise either from locally weak polarization or from
cancellation within a limb sample. A null result therefore does not
imply spatial uniformity. More generally, the method determines how
much spatial mixing is required by the data without assigning that
mixing to a unique emission or propagation mechanism.

The polarization covariance retains information lost in the
scalar bound. For non-zero mean linear polarization, projection
relative to the mean linear-polarization direction separates
variations dominated by polarization degree from those dominated
by polarization angle in the small-variation limit, while the
cross term measures their covariance. The $QV$ and $UV$ terms
further describe co-variation between linear and circular
polarization. Their basis-independent combination
$C^{\rm occ}_{LV}$ measures its amplitude, while $\rho_{LV}$
gives its normalized strength. The anisotropy of the $Q$--$U$
covariance provides complementary directional information,
including when the mean linear polarization vanishes.

These covariance diagnostics distinguish degree--angle,
degree--circular, and angle--circular co-variation and thus
characterize the form, rather than only the amount, of polarization
structure resolved by the occultation. Such correlations may arise
from intrinsic source structure, magnetic geometry, mode balance,
or polarized transfer through a magnetized plasma, but do not by
themselves identify the underlying mechanism. The basis-independent
linear--circular covariance strength is unchanged by a spatially
uniform rotation in the $Q$--$U$ plane, so its frequency dependence
can reveal polarization evolution beyond that produced by a common
Faraday rotation alone.

The inference applies to a stationary source during pure ingress or
pure egress, with each sky position crossed once. A monotonic
intensity curve alone does not guarantee this condition, because
simultaneous covering and uncovering can combine opposite
contributions in the Stokes derivatives. The analysis also assumes
mutually incoherent emission from distinct positions and negligible
diffraction on the relevant scales. Instrumental weighting and
emission from the occulting body must be negligible or treated
consistently. Finite time bins preserve the lower bound for exact
Stokes increments, and nested refinement can only resolve additional
variance. Measurement noise, however, can bias the estimators,
particularly for small intensity increments.
Reliable observational inference therefore requires
the full measurement-noise covariance matrix, including
correlations between successive Stokes increments.

The measurements also provide several complementary diagnostics.
Temporal refinement reveals additional structure along the scan
direction, while variation within a limb sample remains unresolved.
Frequency-resolved measurements trace changes in the scalar and
covariance diagnostics; departures from the behaviour expected under
a common spatially uniform Faraday rotation can indicate intrinsic
spectral evolution or spatially varying propagation. Complementary
limb trajectories can expose structure hidden from a single scan and
strengthen the lower bound on spatial mixing. Solar eclipses provide
a natural setting for radio applications, while occultations and
eclipses of stellar and compact sources offer further possibilities.
Connecting these constraints to physical source parameters requires
the boundary geometry, instrumental response, and an appropriate
emission or propagation model.

Occultation polarimetry thus determines both how much spatial
polarization mixing is required by an unresolved source and what
form of polarization variation is resolved by the scan. It extracts
these constraints directly from full-Stokes light curves while
making explicit the information that remains inaccessible to a
single occultation.

\begin{acknowledgements}
The author thanks V. Kakhiani for valuable discussions on practical 
aspects of polarimetric observations. 
\end{acknowledgements}

\end{document}